\documentclass[prd,aps,twocolumn,floatfix,nofootinbib,preprintnumbers, showpacs,showkeys,superscriptaddress]{revtex4-1}

\usepackage{amsmath,graphicx,color,epsfig}
\usepackage{pstricks}
\usepackage{float}
\usepackage{subfigure}
\usepackage{color}
\usepackage{multirow}

\usepackage[colorlinks=true, linkcolor=blue, citecolor=blue, urlcolor=blue] {hyperref}

\begin{document}

\preprint{PITT PACC-1708}

\title{Renormalization scheme dependence of high-order perturbative QCD predictions}

\author{Yang Ma}
\email{mayangluon@pitt.edu}
\address{Pittsburgh Particle physics, Astrophysics, and Cosmology Center,
Department of Physics and Astronomy, University of Pittsburgh,
3941 O'Hara St., Pittsburgh, PA 15260, USA}

\author{Xing-Gang Wu}
\email{wuxg@cqu.edu.cn}
\address{Department of Physics, Chongqing University, Chongqing 401331, P.R. China}

\date{\today}

\begin{abstract}

Conventionally, one adopts typical momentum flow of a physical observable as the renormalization scale for its perturbative QCD (pQCD) approximant. This simple treatment leads to renormalization scheme-and-scale ambiguities due to the renormalization scheme and scale dependence of the strong coupling and the perturbative coefficients do not exactly cancel at any fixed order. It is believed that those ambiguities will be softened by including more higher-order terms. In the paper, to show how the renormalization scheme dependence changes when more loop terms have been included, we discuss the sensitivity of pQCD prediction on the scheme parameters by using the scheme-dependent $\{\beta_{m \geq 2}\}$-terms. We adopt two four-loop examples, $e^+ e^- \to {\rm hadrons}$ and $\tau$ decays into hadrons, for detailed analysis. Our results show that under the conventional scale setting, by including more-and-more loop terms, the scheme dependence of the pQCD prediction cannot be reduced as efficiently as that of the scale dependence. Thus a proper scale-setting approach should be important to reduce the scheme dependence. We observe that the principle of minimum sensitivity could be such a scale-setting approach, which provides a practical way to achieve optimal scheme and scale by requiring the pQCD approximate be independent to the ``unphysical" theoretical conventions.

\pacs{12.38.Bx, 12.38.Aw, 11.15.Bt}

\end{abstract}

\maketitle

\section{Introduction}

Within the framework of the perturbative quantum chromodynamics (pQCD) theory, a physical observable ($\varrho$) can be expanded up to $n$th order in the strong coupling constant $\alpha_s$ as
\begin{eqnarray}
\varrho_n = \sum_{i=0}^{n} {\cal C}^{\cal R}_{i}(\mu) (a^{\cal R}_s)^{p+i}(\mu), \label{eq.pqcd}
\end{eqnarray}
where ${\cal R}$ stands for the chosen renormalization scheme, $p$ is the power of strong coupling constant associated with the tree-level term, $a^{\cal R}_s(\mu)=\alpha^{\cal R}_s(\mu)/\pi$. Here $\mu$ is the renormalization scale, which is in principle arbitrary but should be within the perturbative region to ensure the reliability of the pQCD expansion. One usually sets the magnitude of $\mu$ to be the same order of the typical momentum flow of the process or some function of the particles momenta to eliminate large logs and achieve a better pQCD convergence, and takes an arbitrary range to estimate the uncertainties in the fixed-order QCD prediction. However, there is no guarantee that the actual fixed-order pQCD prediction lies within the assumed range.

When $n\to\infty$, the infinite series $\varrho_{n\to\infty}$ corresponds to the exact value of the observable and is independent to the choices of renormalization scheme and scale. This is the standard renormalization group (RG) invariance. However this RG invariance is caused by compensation of scheme or scale dependence for all orders; thus, at any finite order, the renormalization scheme and scale dependence from the strong coupling constant and the perturbative coefficients do not exactly cancel, leading to the well-known ambiguities. The fixed-order prediction obtained by using the above mentioned guessed scale depends heavily on the renormalization scheme which is itself arbitrary. Thus, a primary problem for pQCD is how to set the renormalization scale so as to obtain the most accurate fixed-order estimate while satisfying the principles of the renormalization group.

A solution of such ambiguities can be achieved by the exact cancellation of scheme-and-scale dependence at each perturbative order, which can be done by applying the ``principle of maximum conformality'' (PMC)~\cite{Brodsky:2011ta, Brodsky:2012rj, Mojaza:2012mf, Brodsky:2013vpa}; two comprehensive reviews on PMC together with its applications can be found in Refs.\cite{Wu:2013ei, Wu:2014iba}. The PMC is theoretically sound, but its procedures are somewhat complex and depend heavily on how well we know the nonconformal terms of the pQCD series.

It is interesting to know whether there are other approaches which can achieve the same goal. In the paper, we shall concentrate our attention on another solution suggested in the literature, i.e. the ``principle of minimum sensitivity" (PMS)~\cite{Stevenson:1981vj}, which sets the optimal renormalization scale and renormalization scheme by directly requiring the slope of the pQCD approximant over the scheme and scale changes vanish. The PMS, which does not satisfy the RG properties of symmetry, reflexivity, and transitivity~\cite{Brodsky:2012ms}, gives an incorrect prediction in the low-energy region~\cite{Kramer:1987dd}. It could be a practical way to achieve a reliable pQCD prediction when enough higher-order terms are included.

Before we go any further, we first give some explanations on the RG equation, or the so-called $\beta^{\cal R}$-function, which is important for all scale-setting approaches. Generally, the scale running behavior of $\alpha^{\cal R}_s(\mu)$ is controlled by the following $\beta^{\cal R}$-function
\begin{eqnarray}
\beta^{\cal R}=\mu^2\frac{\partial}{\partial\mu^2} \left(\frac{\alpha^{\cal R}_s(\mu)}{4\pi}\right) =-\sum_{i=0}^{\infty}\beta^{\cal R}_{i}\left(\frac{\alpha^{\cal R}_s(\mu)}{4\pi}\right)^{i+2}. \label{eq.beta}
\end{eqnarray}
Using the decoupling theorem, only the first two loop terms $\beta_{0}$ and $\beta_{1}$ are scheme independent~\cite{Politzer:1973fx, Callan:1970yg, Khriplovich:1969aa, Caswell:1974gg, Jones:1974mm, Egorian:1978zx}, while the high-order $\{\beta^{\cal R}_{m\geq 2}\}$-terms are scheme dependent, which are now calculated up to the five-loop level within the conventional $\overline{\rm MS}$ scheme~\cite{Tarasov:1980au, Larin:1993tp, vanRitbergen:1997va, Chetyrkin:2004mf, Czakon:2004bu, Baikov:2016tgj}.

Following the idea of effective coupling approach~\cite{Grunberg:1982fw}, any physical observable can be equivalently expressed by an effective coupling which satisfies the similar RG equation. Thus the $\{\beta^{\cal R}_{m\geq 2}\}$-terms in the pQCD approximant of a physical observable can be inversely adopted to get the correct running behavior of the coupling constant. For the PMC scale-setting approach, if one can tick out which $\{\beta^{\cal R}_{i}\}$-terms pertain to which perturbative order, one can achieve the correct running behavior by using the RG equation and set the correct scale for the strong coupling at this particular order. However if it is hard to do such $\{\beta^{\cal R}_{m\geq 2}\}$-terms distribution, it is helpful to find a proper approach to deal with them as a whole.

Different renormalization schemes lead to different $\{\beta^{\cal R}_{m\geq 2}\}$-terms, thus those $\{\beta^{\cal R}_{m\geq 2}\}$-terms can be inversely adopted to characterize a renormalization scheme. For the purpose, Refs.\cite{Stevenson:1981vj, Lu:1992nt} extended the RG equation to the extended RG equations to incorporate both the scale-running and scheme-running behaviors of the coupling, especially via this way the strong coupling at different scales and schemes can be reliably related via a continuous way, since along the evolution trajectory described by the extended RG equations, no dissimilar scales or schemes are involved.

\section{Transformation of pQCD prediction from one scheme to another scheme}

As a practical treatment, one suggests that we can get optimal scheme and scale by requiring the pQCD approximate at any fixed order be independent of the ``unphysical" theoretical conventions. This is also the key idea of PMS, which indicates that all the scheme-and-scale dependence of a fixed-order prediction are treated as the negligible high-order effect,
\begin{equation}
\partial \varrho_n /\partial {\rm (RS)}={\cal O}(a_s^{p+n+1})\sim 0,
\end{equation}
where ${\rm RS}$ stands for the scheme or scale parameters. Equivalently, it requires the fixed-order approximant $\varrho_n$ to satisfy the local RG invariance~\cite{Wu:2014iba, Ma:2014oba}
\begin{eqnarray}
\frac{\partial \varrho_n}{\partial \tau} &=& 0,  \label{eq.PMSscale}\\
\frac{\partial \varrho_n}{\partial \beta^{\cal R}_m} &=& 0, \;\; (2\leq m \leq n) \label{eq.PMSscheme}
\end{eqnarray}
where $\tau=\ln(\mu^2/\tilde\Lambda^2_{\rm QCD})$ with the alternated asymptotic QCD scale $\tilde\Lambda_{\rm QCD}= \left({\beta_1}/ {\beta_0^2}\right)^{-\beta_1/2\beta_0^2} \Lambda^{\cal R}_{\rm QCD}$.

The integration constants of those differential equations, i.e. Eqs. (\ref{eq.PMSscale}) and (\ref{eq.PMSscheme}), are scheme-and-scale independent RG invariants. For example, up to the ${\rm N^3LO}$ level, there are three RG invariants,
\begin{eqnarray}
\rho_1 =&& {1\over 4} p \beta_0 \tau-{\cal C}^{'\cal R}_{1}, \label{eq.rho1} \\
\rho_2 =&& {\cal C}^{'\cal R}_2-\frac{(1+p) ({\cal C}^{'\cal R}_1)^2}{2 p}-\frac{\beta _1 {\cal C}^{'\cal R}_1}{4 \beta _0}+\frac{p \beta^{\cal R}_2}{16 \beta _0} , \label{eq.rho2}\\
\rho_3 =&& 2 {\cal C}^{'\cal R}_3+\frac{({\cal C}^{'\cal R}_1)^2 \beta _1}{4 p \beta _0}-\frac{{\cal C}^{'\cal R}_1 \beta^{\cal R}_2}{8 \beta _0}+\frac{p \beta^{\cal R}_3}{64 \beta _0} \nonumber \\
&&+\frac{2 (1+p) (2+p) ({\cal C}^{'\cal R}_1)^3}{3 p^2}-\frac{2 (2+p) {\cal C}^{'\cal R}_1 {\cal C}^{'\cal R}_2}{p}, \label{eq.rho3}
\end{eqnarray}
where ${\cal C}'_i={\cal C}_i/{\cal C}_0$. In combination with the known RG invariants, the local RG equations (\ref{eq.PMSscale}) and (\ref{eq.PMSscheme}), and the solution of the RG equation (\ref{eq.beta}) up to the same order as the pQCD approximant, we are ready to derive all the wanted optimal parameters. At high-orders, it can be done numerically by using the ``spiraling" method~\cite{Mattingly:1992ud, Mattingly:1993ej}.

It is noted that those RG invariants are also helpful for transforming the pQCD approximant $\varrho_n$ under the ${\cal R}$ scheme to the one under any other scheme (labeled as the ${\cal S}$ scheme). More explicitly, this transformation can be achieved by applying the following two transformations simultaneously:
\begin{equation}
a^{\cal R}_s \to a^{\cal S}_s\;\; {\rm and}\;\; {\cal C}_i^{\cal R} \to {\cal C}_i^{\cal S}.
\end{equation}
The coupling constant $a^{\cal S}_s$ can be derived from $a^{\cal R}_s$ by using the extended RG equations, and the scheme-dependent $\beta^{\cal S}_{i\geq2}$-terms which determine $a^{\cal S}_s$ scale running behavior can be achieved by using the relation,
\begin{equation}
\beta^{\cal S}(a^{\cal S}_s) = \left({{\partial a^{\cal S}_s}}/{{\partial a^{\cal R}_s}}\right) {\beta^{\cal R}}({a^{\cal R}_s}).
\end{equation}
The coefficients ${\cal C}_i^{\cal S}$ can be obtained from the coefficients ${\cal C}_i^{\cal R}$ by using the RG invariants $\rho_i$, e.g. up to the N$^3$LO level, we have
\begin{eqnarray}
{\cal C}_1^{\cal S}=&&{\cal C}_1^{\cal R}, \label{eq.coetrans1}\\
{\cal C}_2^{\cal S}=&&{\cal C}_2^{\cal R}+\frac{p}{16 \beta _0}(\beta _2^{\cal R}-\beta_2^{\cal S}), \label{eq.coetrans2}\\
{\cal C}_3^{\cal S}=&&{\cal C}_3^{\cal R}+\frac{p+1}{16 \beta _0}{\cal C}_1^{\cal R}(\beta _2^{\cal R}-\beta_2^{\cal S})  +\frac{p}{128}(\beta _3^{\cal R}-\beta_3^{\cal S}). \label{eq.coetrans3}
\end{eqnarray}

So far we have explained how to transform the pQCD predictions from one scheme to another scheme and have completed the description of the PMS calculation technology. In the following, we shall take two four-loop examples to show the scheme dependence of a pQCD approximant under conventional and PMS scale-setting approaches, respectively.

\section{Comparisons of the pQCD predictions under different schemes}

To do the numerical calculation, the initial scheme is taken as the usual ${\overline {\rm MS}}$ scheme~\cite{Bardeen:1978yd}, and the QCD parameter $\Lambda_{\rm QCD}$ is fixed by $\alpha_s(M_Z)=0.1181$ \cite{Olive:2016xmw}. \\

\noindent{\bf Example I : $e^+e^- \to$ {hadrons}.} The annihilation of an electron and positron into hadrons provides one of the most precise platforms for testing the running behavior of the strong coupling constant. Its characteristic parameter is the $R$-ratio, whose definition is
\begin{eqnarray}
R_{e^+e^-}(Q)&=&\frac{\sigma\left(e^+e^-\rightarrow {\rm hadrons} \right)}{\sigma\left(e^+e^-\rightarrow \mu^+\mu^-\right)}\nonumber\\
&=& 3\sum_q e_q^2\left[1+R(Q)\right], \label{eq.Re+e-}
\end{eqnarray}
where $Q$ stands for the $e^+e^-$ collision energy at which the $R$-ratio is measured. The pQCD approximant of $R(Q)$ up to the N$^{n}$LO level under the $\overline{\rm MS}$ scheme reads
\begin{equation}
R_n(Q,\mu_0)=\sum_{i=0}^{n} {\cal C}^{\overline{\rm MS}}_{i}(Q,\mu_0) (a^{\overline{\rm MS}}_s)^{i+1}(\mu_0),
\end{equation}
where $\mu_0$ stands for an arbitrary initial renormalization scale. If setting $\mu_0=Q$, the coefficients ${\cal C}^{\overline{\rm MS}}_{i}(Q,Q)$ up to fourth order can be read from Ref.\cite{Baikov:2012zn}. For any other choice of $\mu_0$, we will use RG equation to get the coefficients from ${\cal C}^{\overline{\rm MS}}_{i}(Q,Q)$.

\begin{figure}[htb]
\centering
\includegraphics[width=0.5 \textwidth]{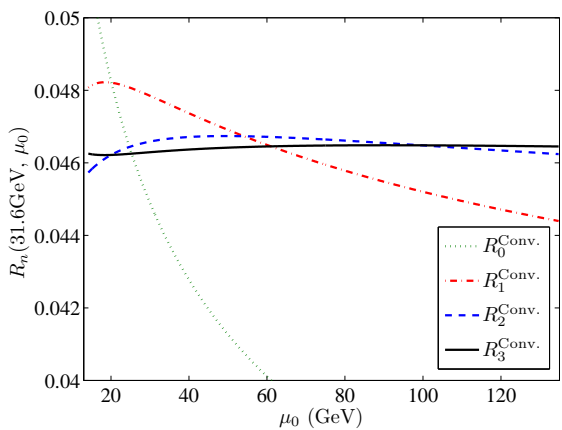}
\caption{The pQCD approximant $R_{n}(Q=31.6 {\rm GeV},\mu_{0})$ up to four-loop level versus the initial scale $\mu_0$. The dotted, the dash-dotted, the dashed and the solid lines are for $R_0$, $R_1$, $R_2$ and $R_3$, respectively. } \label{fig.Rsmu0}
\end{figure}

As a reference, we present the recalculated conventional $\mu_0$-dependence of $R_{n}(Q,\mu_0)$ under the $\overline{\rm MS}$ scheme in Figure \ref{fig.Rsmu0}, where $Q=31.6$ GeV. A combination of data from $e^+ e^-$ colliders gives $R(31.6GeV)=0.0527\pm0.0050$~\cite{Marshall:1988ri}. Such $\mu_0$-dependence as shown by Figure \ref{fig.Rsmu0} is rightly the renormalization scale dependence for conventional scale setting, since $\mu\equiv\mu_0$ for conventional scale setting. Conventionally, people would like to choose $\mu_0=Q$ and vary it in range $[Q/2,2Q]$ to estimate the uncertainty due to scale ambiguity. Figure \ref{fig.Rsmu0} shows that under conventional scale setting, the low-order results, e.g. $R_0$ and $R_1$, depend strongly on the renormalization scale, which becomes weaker-and-weaker when more-and-more high-order terms have been included. This agrees with the conventional wisdom on the perturbation theory that one would get a desirable renormalization scale invariant result by finishing enough high-order calculations.

Next, we investigate the renormalization scheme dependence of the pQCD predictions at N$^{2}$LO and beyond. We will also show how the PMS prediction changes with different choice of initial schemes as a comparison. Since the PMS prediction is $\mu_0$-independent \cite{Ma:2014oba}, it is safe to fix $\mu_0=Q$ in the following discussions.

At the N$^2$LO level and higher, the scheme dependence of the pQCD prediction could be equivalently represented by a group of scheme-dependent $\{\beta_{m\geq2}\}$-terms, since the $\{\beta_{m\geq2}\}$-terms are specific for a specific scheme. At the N$^n$LO level, the number of $\{\beta_{m\geq2}\}$-terms is $(n-1)$, which characterize the scheme independence of the PMS prediction via the local version of RG equations (\ref{eq.PMSscheme}). It is worthy to point out that, unlike the renormalization scale which should be varied in a region that is not too far away from a character value $Q$ of a given process, there is no such constraint for $\{\beta_{m\geq2}\}$-terms. In principle one could choose any value for $\{\beta_{m\geq2}\}$-terms and define a $\cal R$ scheme for a given fixed order pQCD calculation, as long as $\beta^{\cal R} (\alpha_s(Q)) < 0$, i.e. the asymptotic freedom is satisfied. Following the idea,
\begin{enumerate}
    \item We use $R_{n,m}(Q,\beta_{m})$ to replace $R_{n}$ to show explicitly how the scheme-dependent $\beta_m$-term affects the ${\rm N}^n$LO prediction $R_n$.
    \item When discussing the scheme error from the $\beta_{m}$-term, the other involved $\{\beta_{j\neq m}\}$-terms shall be fixed to be their $\overline{\rm MS}$-scheme values.
    \item We adopt a broad range for the allowable $\{\beta_{m}\}$-terms to numerically discuss the scheme dependence of a N$^n$LO prediction, e.g. $\beta_{m}\in [-50\beta_{m}^{\over {\rm MS}},+50\beta_{m}^{\over {\rm MS}}]$ for $m\in[2,n]$.
\end{enumerate}

\begin{figure}[htb]
\centering
\includegraphics[width=0.5 \textwidth]{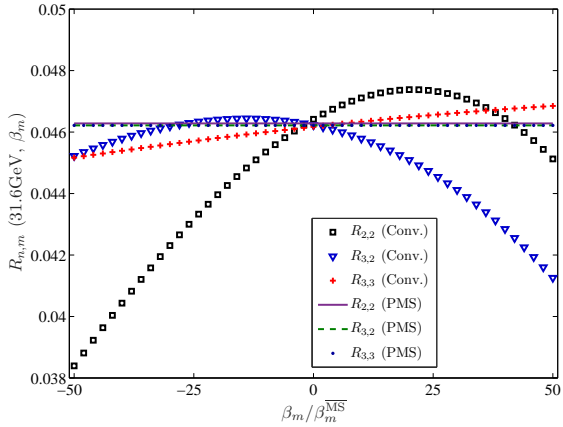}
\caption{The pQCD predictions for $R_{n,m}(Q,\beta_m)$ versus the $\beta_{m\geq2}$-term at the ${\rm N}^2$LO ($n=2$) and ${\rm N}^3$LO ($n=3$) levels respectively. Q=31.6 {\rm GeV}. As a comparison, both the results for conventional scale setting and PMS are presented. } \label{fig.Rsbeta1}
\end{figure}

The dependence of the ratio $R_{n,m}(Q,\beta_{m})$ on $\beta_{m}$ with $m\in[2,n]$ is presented in Figure \ref{fig.Rsbeta1}, where $n=[2,3]$ and $Q=31.6{\rm GeV}$. Here $n=2$ and $n=3$ stand for the ${\rm N}^2$LO and ${\rm N}^3$LO pQCD predictions, respectively. As a comparison, both the results for conventional scale setting and PMS are presented. Under conventional scale setting,
\begin{itemize}
    \item Figure \ref{fig.Rsbeta1} shows the scheme dependence of the pQCD calculation follows the perturbative nature, which drops down when more perturbative terms are included.

    \item Comparing with Figure \ref{fig.Rsmu0}, Figure \ref{fig.Rsbeta1} shows the magnitude of the scheme dependence could be much larger than that of the scale dependence:
    \begin{itemize}
        \item At the ${\rm N}^2$LO level, the scale dependence is $\sim 2\%$ for $\mu_0\in[Q/2,2Q]$, while the scheme dependence is $\sim 4\%$ for $\beta_{2}\in [-10\beta_{2}^{\over {\rm MS}}, +10\beta_{2}^{\over {\rm MS}}]$ and $\sim 19\%$ for $\beta_{2}\in [-50\beta_{2}^{\over {\rm MS}},+50\beta_{2}^{\over {\rm MS}}]$. Figure \ref{fig.Rsbeta1} shows that $R_{2,2}(Q,\beta_2)$ shall first increase and then decrease with the increment of $\beta_{2}$.
        \item At the ${\rm N}^3$LO level, the scale dependence reduces to be $\sim0.2\%$, while the scheme dependence is still about $7\%$ in which an extra $\sim 4\%$ error comes from the $\beta_3$-term within the region of $[-50\beta_{3}^{\over {\rm MS}},+50\beta_{3}^{\over {\rm MS}}]$. Figure \ref{fig.Rsbeta1} shows $R_{3,2}(Q,\beta_2)$ shall first increase and then decrease with the increment of $\beta_{2}$, and $R_{3,3}(Q,\beta_3)$ shall monotonously increase with the increment of $\beta_{3}$.
    \end{itemize}
\end{itemize}
The PMS determines the optimal scheme and scale by requiring the slope of the pQCD prediction to vanish. Figure \ref{fig.Rsbeta1} shows the PMS prediction is stable over the scheme changes. That is, the flat lines for $R_{2,2}(Q,\beta_2)|_{\rm PMS}$, $R_{3,2}(Q,\beta_2)|_{\rm PMS}$ and $R_{3,3}(Q,\beta_3)|_{\rm PMS}$ indicate that the PMS predictions are scheme independent at each order.

\begin{figure}[htb]
\centering
\includegraphics[width=0.5 \textwidth]{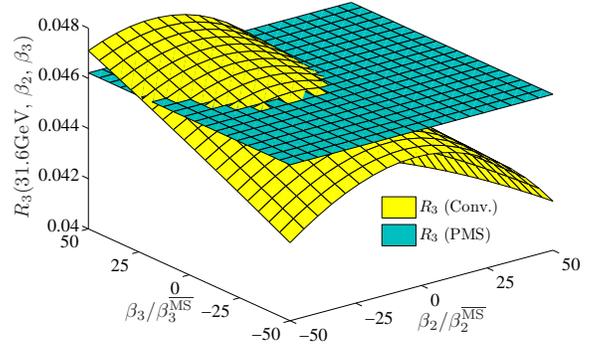}
\caption{Comparison of the combined $\{\beta_{2},\beta_{3}\}$-dependence for the ${\rm N}^{3}$LO prediction $R_{3}$ under the conventional (Conv.) scale setting and PMS, respectively. Q=31.6 {\rm GeV}. } \label{fig.Rsbeta2}
\end{figure}

Figure \ref{fig.Rsbeta2} shows the combined $\{\beta_{2},\beta_{3}\}$-dependence for the ${\rm N}^{3}$LO prediction $R_{3}$, in which the scheme-dependent $\beta_{2}$ and $\beta_{3}$ terms change simultaneously within the region of $[-50\beta_{2}^{\over {\rm MS}},+50\beta_{2}^{\over {\rm MS}}]$ and $[-50\beta_{3}^{\over {\rm MS}},+50\beta_{3}^{\over {\rm MS}}]$. The flat plane confirms the scheme independence of the PMS prediction over the changes of $\{\beta_2,\beta_3\}$, thus by using the scheme equations (\ref{eq.PMSscheme}), one cannot only achieve the most stable pQCD prediction around the optimal point (determined by the optimal scheme and the optimal scale) but also achieve the scheme-independent prediction over the choices of $\{\beta_m\}$-terms, or equivalently over different choices of the initial renormalization scheme.

\begin{table}[htb]
\centering
\begin{tabular}{ccccccc}
\hline
                            & $R_1$   & $R_2$   & $R_3$   & $\kappa_1$ & $\kappa_2$ & $\kappa_3$ \\ \hline
$\overline {\rm MS}$ (Conv.) & 0.04765 & 0.04650 & 0.04619 & -          & -          & -          \\
MOM (Conv.)                  & 0.04810 & 0.04604 & 0.04608 & 0.9\%     & 1.0\%     & 0.2\%    \\
V (Conv.)                    & 0.04801 & 0.04653 & 0.04587 & 0.8\%     & 0.1\%     & 0.7\%    \\
\hline
\end{tabular}
\caption{Numerical results for $R_n$ and $\kappa_n$ with various QCD loop corrections under the conventional scale setting. $\mu_0=Q$ and $Q=31.6$ GeV. Three renormalization schemes, $\overline{\rm MS}$, MOM and V, are adopted for a comparison.}
\label{tab.Rs}
\end{table}

\begin{table}[htb]
\centering
\begin{tabular}{cccc}
\hline
                                   & $n_f=3$ & $n_f=4$ & $n_f=5$  \\ \hline
$\beta^{\overline{\rm MS}}_2$ & 643.833 & 406.352 & 180.907  \\
$\beta^{\rm MOM}_2$                    & 1338.77 & 849.069 & 398.132  \\
$\beta^{\rm V}_2$                      & 2174.01 & 1574.11 & 1015.95  \\
\hline
$\beta^{\overline{\rm MS}}_3$ & 12090.4 & 8035.19 & 4826.16  \\
$\beta^{\rm MOM}_3$                    & 41157.4 & 27094.6 & 15622.9  \\
$\beta^{\rm V}_3$                      & 10537.9 & 2355.74 & -4529.47 \\ \hline
\end{tabular}
\caption{Numerical values for $\beta_2$ and $\beta_3$ under the three renormalization schemes, $\overline{\rm MS}$, MOM and V, respectively. }
\label{tab.beta}
\end{table}

To be more specific, we present the numerical results of $R_n$ under three usually adopted schemes, i.e. $\overline{\rm MS}$ scheme, MOM scheme~\cite{Celmaster:1979dm} and V scheme~\cite{Peter:1996ig}, in Table \ref{tab.Rs}. Typical values for $\beta_2$ and $\beta_3$ of those renormalization schemes are presented in Table \ref{tab.beta}. Here, the result for the MOM scheme is obtained by using the Landau gauge and following the method of Ref.\cite{Chetyrkin:2000dq}; the result for the V scheme is consistent with Ref.\cite{Kataev:2015yha}. For $\mu_0=Q=31.6$ GeV, which corresponds to $n_f=5$, we have
\begin{eqnarray}
\frac{\beta^{\rm MOM}_2}{\beta^{\overline{\rm MS}}_2}&\simeq&2.2, \;\; \frac{\beta^{\rm V}_2}{\beta^{\overline{\rm MS}}_2}\simeq5.6, \\
\frac{\beta^{\rm MOM}_3}{\beta^{\overline{\rm MS}}_3}&\simeq&3.2, \;\; \frac{\beta^{\rm V}_3}{\beta^{\overline{\rm MS}}_3}\simeq-0.9.
\end{eqnarray}
In Table \ref{tab.Rs}, to show the scheme dependence, we have defined the ratio $\kappa^{\cal R}_n$ for a specific scheme ${\cal R}$:
\begin{eqnarray}
\kappa_n^{\cal R}=\left | \frac{R_n^{\cal R}-R_n^{\overline {\rm MS}}}{R_n^{\overline {\rm MS}}} \right |,  \;\;(n\in[1,3]).
\end{eqnarray}
Under conventional scale setting, by setting $\mu_0=Q$, we observe that $\kappa^{\rm MOM,V}_1 \sim 1\%$ for $Q=31.6$ GeV, which generally drops down as more loop terms come into contribution. However, such a shrink tendency heavily depends on the cancellations among different perturbative orders which could be accidental. For example, $\kappa^{\rm V}_3 \sim \kappa^{\rm V}_1$ and $\kappa^{\rm V}_3 \gg \kappa^{\rm V}_2$. This anomaly can be qualitatively explained by the ascending trends of $R_{n,m}(Q,\beta_m)$ versus the $\{\beta_{m\geq2}\}$-terms as shown by Figure \ref{fig.Rsbeta1}, $R_{3,2}(Q,\beta_2)$ first increases and then decreases with the increment of $\beta_{2}$, and $R_{3,3}(Q,\beta_3)$ monotonously increases with the increment of $\beta_{3}$. Thus the fact of ${\beta^{\rm V}_2}>{\beta^{\overline{\rm MS}}_2}$ and ${\beta^{\rm V}_3}<{\beta^{\overline{\rm MS}}_3}$ indicates there is no cancellation between ${\beta^{\rm V}_2}$-terms and ${\beta^{\rm V}_3}$-terms at the N$^3$LO level, leading to a larger difference between $R^{\rm V}_3=0.04587$ and $R^{\overline{\rm MS}}_3=0.04619$.

On the other hand, being consistent with previous observations, after applying PMS scale setting, the scheme dependence is eliminated and all three schemes lead to the same predictions~\footnote{The four-loop pQCD prediction before and after applying PMS scale setting gives $R(31.6{\rm GeV})\sim 0.046$, which is smaller than the measured value $[0.0477, 0.0557]$~\cite{Marshall:1988ri}. A precise determination of $\Lambda_{\rm QCD}$ by using the $e^+ e^-$ data along shall be helpful for explanation of such difference, which is in preparation.}
\begin{equation}
R_1=0.04868,\;\; R_2=0.04628,\;\; R_3=0.04622.
\end{equation}

\noindent{\bf Example II : $\tau$ decays into hadrons.} The $\tau$ decays into hadrons is another important platform to test the pQCD theory, whose characteristic parameter is the following ratio,
\begin{eqnarray}
R_{\tau}(M_{\tau}) &=&\frac{\Gamma(\tau\rightarrow\nu_\tau+\rm{hadrons})} {\Gamma(\tau\rightarrow\nu_\tau+e^-\bar\nu_e)} \nonumber\\
&=&3(|V_{ud}|^2+|V_{us}|^2) \left[1+r^{\tau}(M_{\tau})\right],
\end{eqnarray}
where the $\tau$-lepton mass $M_{\tau}=1.777$ GeV~\cite{Olive:2016xmw} and the Cabbibo-Kobayashi-Maskawa matrix elements $V_{ud}$ and $V_{us}$ satisfy the relation, $3(|V_{ud}|^2+|V_{us}|^2)\approx 3$. The pQCD approximant of $r^{\tau}(M_{\tau})$ up to the N$^{n}$LO level under the $\overline{\rm MS}$ scheme reads
\begin{equation}
r^{\tau}_n(M_{\tau},\mu_0)=\sum_{i=0}^{n} {\cal C}^{'\overline{\rm MS}}_{i}(M_{\tau},\mu_0) (a^{\overline{\rm MS}}_s)^{i+1}(\mu_0).
\end{equation}
The perturbative coefficients up to fourth order at any scale $\mu_0$ can be derived from the ones in Ref.\cite{Baikov:2008jh}.

\begin{figure}[htb]
\centering
\includegraphics[width=0.5 \textwidth]{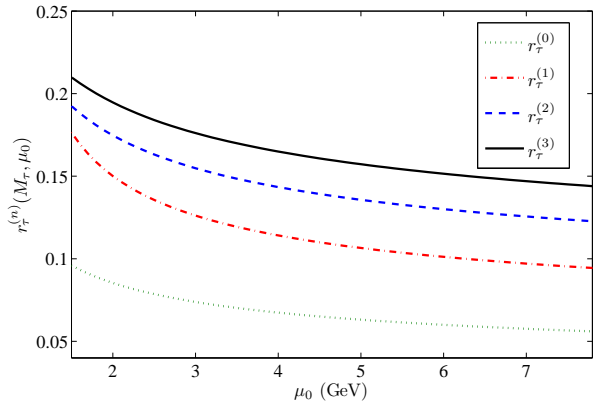}
\caption{The pQCD approximant $r^{\tau}_n(M_{\tau},\mu_0)$ up to four-loop level versus the initial scale $\mu_0$. The dotted, the dash-dotted, the dashed, and the solid lines are for $r^{\tau}_0$, $r^{\tau}_1$, $r^{\tau}_2$ and $r^{\tau}_3$, respectively.}
\label{fig.Rtaumu1}
\end{figure}

Using the above formulae, we calculate the initial scale dependence of $r^{\tau}_n(M_{\tau},\mu_0)$ and present it in Figure \ref{fig.Rtaumu1}, which shows a much larger scale dependence than that of $R_n(Q,\mu_0)$. The reason for this larger scale dependence even up to the four-loop level is due to the poorer pQCD convergence of the conventional pQCD series, which is caused by the divergent renormalon terms in combination of a somewhat larger $\alpha_s$ value at the lower scale around $M_\tau$ that is not far from $\Lambda_{\rm QCD}$.  We observe that at the four-loop level, under conventional scale setting, the scale dependence is still about $25\%$ for $\mu_0\in[M_{\tau}/2,2M_{\tau}]$. Thus an even higher order calculation is necessary to further suppress the conventional scale uncertainty. After applying PMS scale setting, the PMS prediction on $R_n(Q,\mu_0)$ is independent to the choice of $\mu_0$, and for convenience, we set $\mu_0=M_\tau$ to do the following discussion.

\begin{figure}[htb]
\centering
\includegraphics[width=0.5 \textwidth]{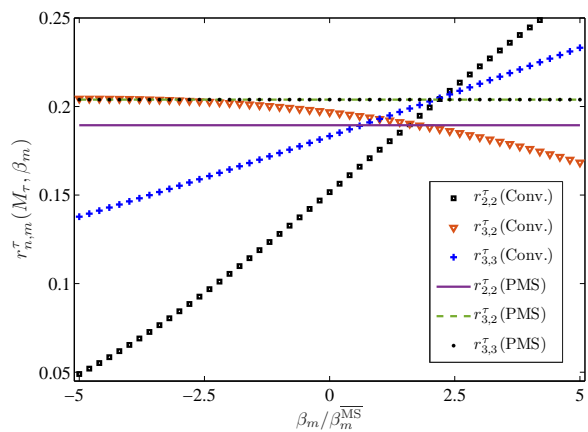}
\caption{The pQCD predictions for $r^{\tau}_n(M_{\tau},\beta_m)$ versus the $\beta_{m\geq2}$-term at the ${\rm N}^2$LO ($n=2$) and ${\rm N}^3$LO ($n=3$) levels respectively. As a comparison, both the results for conventional scale setting and PMS are presented. }
\label{fig.Rtaubeta1}
\end{figure}

\begin{figure}[htb]
\centering
\includegraphics[width=0.5 \textwidth]{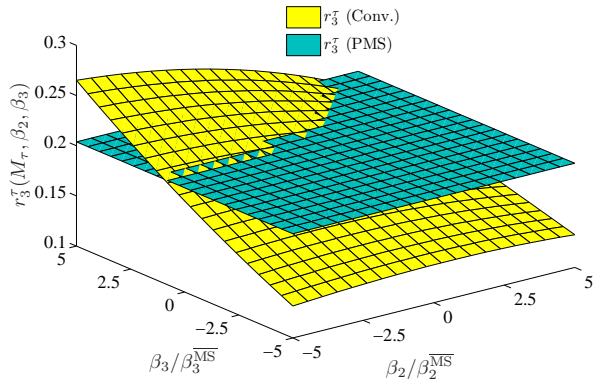}
\caption{Comparison of the combined $\{\beta_{2},\beta_{3}\}$-dependence for the ${\rm N}^{3}$LO prediction $r^{\tau}_3$ under the conventional (Conv.) scale setting and PMS, respectively. }
\label{fig.Rtaubeta2}
\end{figure}

Similarly, we use $r^{\tau}_{n,m}(M_{\tau},\beta_m)$ instead of $r^{\tau}_{n}$ to discuss the scheme dependence. The magnitude of the conventional scheme dependence is large, so we adopt a smaller region, $\beta_m\in [-5\beta_{m}^{\over {\rm MS}},+5\beta_{m}^{\over {\rm MS}}]$ for the discussion. The results for the $\beta_m$ dependence of $r^{\tau}_{n,m}(M_{\tau},\beta_m)$ before and after PMS scale setting are presented in Figures \ref{fig.Rtaubeta1} and \ref{fig.Rtaubeta2}, where $n\in[2,3]$.

To compare with Figure \ref{fig.Rsbeta1}, Figure \ref{fig.Rtaubeta1} shows that $r^{\tau}_{n,m}(M_{\tau},\beta_m)$ has a much heavier $\beta$-dependence than that of $R_{n,m}(Q,\beta_m)$, which increases (decreases) monotonously with the increment of $\beta_2$ ($\beta_3$). At the ${\rm N}^3$LO level, the conventional scheme uncertainty is about $30\%$ for $\beta_2\in [-5\beta_{2}^{\over {\rm MS}},+5\beta_{2}^{\over {\rm MS}}]$ and $\beta_3\in [-5\beta_{3}^{\over {\rm MS}},+5\beta_{3}^{\over {\rm MS}}]$. The flat lines in Figure \ref{fig.Rtaubeta1} and the flat plane in Figure \ref{fig.Rtaubeta2} show the PMS predictions are stable over the scheme changes.

\begin{table}[htb]
\centering
\begin{tabular}{ccccccc}
\hline
                            & $r^{\tau}_1$ & $r^{\tau}_2$ & $r^{\tau}_3$ & $\kappa^{\tau}_1$ & $\kappa^{\tau}_2$ & $\kappa^{\tau}_3$ \\
\hline
$\overline {\rm MS}$(Conv.) & 0.1544       & 0.1755       & 0.1930       & -                 & -                 & -                 \\
MOM(Conv.)                  & 0.2389       & 0.2581       & 0.1423       & 55\%            & 47\%            & 26\%            \\
V(Conv.)                    & 0.1975       & 0.2773       & 0.1719       & 28\%            & 58\%            & 11\%           \\
\hline
\end{tabular}
\caption{Numerical results for $r^{\tau}_n$ and $\kappa^{\tau}_n$ with various QCD loop corrections under the conventional scale setting. $\mu_0=M_\tau$. Three renormalization schemes, $\overline{\rm MS}$, MOM and V, are adopted for a comparison.}
\label{tab.Rtau}
\end{table}

Typical results of $r^{\tau}_n$ and $\kappa^{\tau}_n$ under the $\overline{\rm MS}$ scheme, the MOM scheme and the V scheme are presented in Table \ref{tab.Rtau}. On the other hand, being consistent with the previous observations, after applying PMS scale setting, the scheme dependence is eliminated and all three schemes lead to the same predictions
\begin{equation}
r^{\tau}_1=0.3238,\;\; r^{\tau}_2=0.1894,\;\; r^{\tau}_3=0.2039.
\end{equation}

\section{Summary}

We have investigated the renormalization scheme dependence of high-order pQCD predictions via studying the sensitivity of the pQCD calculations on the renormalization scheme parameters $\{\beta^{\cal R}_{m\ge2}\}$. For a fixed-order prediction, by simply using a guessed scale as done by conventional scale setting, there are renormalization scheme and scale ambiguities. Those ambiguities could be softened by finishing more and more loop terms.

We observe that, different to the scale dependence, because new scheme-dependent $\beta$-terms emerge at higher orders which introduce new scheme dependence into the prediction, the scheme dependence tends to be dropped down much slower. For example, the scale dependence of the N$^{2}$LO prediction $R_2$ is around $2\%$ for $\mu\in[Q/2,2Q]$ with $Q=31.6$ GeV, while the scheme dependence is still $\sim 19\%$ for $\beta_{2}\in[-50\beta_{2}^{\over {\rm MS}},+50\beta_{2}^{\over {\rm MS}}]$. Moving to the ${\rm N}^3$LO level, the scale dependence of $R_3$ reduces to be less than $0.2\%$ for $\mu\in[Q/2,2Q]$ with $Q=31.6$ GeV and the scheme dependence is $\sim 7\%$ for $\beta_{2}\in[-50\beta_{2}^{\over {\rm MS}},+50\beta_{2}^{\over {\rm MS}}]$ and $\beta_{3}\in[-50\beta_{3}^{\over {\rm MS}},+50\beta_{3}^{\over {\rm MS}}]$; meanwhile, the scale dependence of $r^{\tau}_3$ is $\sim 25\%$ for $\mu\in[M_\tau/2,2M_\tau]$ and the scheme dependence is about $30\%$ for $\beta_{2}\in[-5\beta_{2}^{\over {\rm MS}},+5\beta_{2}^{\over {\rm MS}}]$ and $\beta_{3}\in[-5\beta_{3}^{\over {\rm MS}},+5\beta_{3}^{\over {\rm MS}}]$.

We have adopted three schemes, the $\overline{\rm MS}$ scheme, the MOM scheme and the V scheme, to show how the scheme dependence changes when more loop terms are included. Tables \ref{tab.Rs} and \ref{tab.Rtau} show $\kappa^{\rm V}_3 \sim \kappa^{\rm V}_1$, $\kappa^{\rm V}_3 \gg \kappa^{\rm V}_2$ for $R$ and $\kappa^{\rm MOM}_2 \sim \kappa^{\rm MOM}_1$, $\kappa^{\rm V}_2 \gg \kappa^{\rm V}_1$ for $r^\tau$, reflecting that those high-order terms cannot effectively suppress the scheme uncertainty as far as they do for the scale uncertainty.

As a summary, it has been found that the elimination of the scheme dependence is as important as the elimination of the scale dependence. It is important to find a proper approach to deal with both issues simultaneously. The PMS treats the fixed-order pQCD prediction as the exact prediction of the physical observable, which breaks the RG invariance. Figures \ref{fig.Rsbeta1}, \ref{fig.Rsbeta2}, \ref{fig.Rtaubeta1}, and \ref{fig.Rtaubeta2} show that by applying the PMS, one can achieve scheme-and-scale independent predictions with the help of RG invariants such as those of Eqs.(\ref{eq.rho1},\ref{eq.rho2},\ref{eq.rho3}). Even though the PMS cannot offer correct lower-order predictions~\cite{Ma:2014oba}, it shall provide reliable prediction when enough higher-order terms are included. Thus, in certain cases when the pQCD series has a good convergence, the PMS could be a practical approach to soften the renormalization scheme and scale ambiguities for high-order pQCD predictions.  \\

\noindent{\bf Acknowledgement}: We thank Hua-Yong Han and Xu-Chang Zheng for helpful discussions. This work was supported in part by National Natural Science Foundation of China under Grant No.11625520. PITT PACC-1708.

\end{document}